\documentclass[12pt,preprint]{aastex}
\RequirePackage{xspace}

\usepackage[T1]{fontenc} 
\usepackage[english]{babel}
\usepackage{graphicx}
\usepackage[dvips]{geometry}

\RequirePackage{xspace}
\RequirePackage{amsmath}
\RequirePackage{amsfonts}
\RequirePackage{amssymb}

\def\ifundefined#1{\expandafter\ifx\csname#1\endcsname\relax}

\def\la{\mathrel{\hbox{\rlap{\hbox{\lower4pt\hbox{$\sim$}}}\hbox{$<$}}}}
\def\ga{\mathrel{\hbox{\rlap{\hbox{\lower4pt\hbox{$\sim$}}}\hbox{$>$}}}}

\newcommand{\be}{\begin{equation}}
\newcommand{\ee}{\end{equation}}

\newcommand{\bea}{\begin{eqnarray}}
\newcommand{\eea}{\end{eqnarray}}

\ifundefined{ensuremath}\def\ensuremath#1{\relax\ifmmode{#1}}
\else${#1}$\fi\else\relax\fi
\ifundefined{nuc}\def\nuc#1#2{\relax\ifmmode{}^{#1}{\protect\mathrm{#2}}
\else${}^{#1}$#2\fi}\else\relax\fi

\newcommand{\kmps}{\ensuremath{\mathrm{km}~\mathrm{s}^{-1}}\xspace}

\newcommand{\msol}{\ensuremath{{\mathrm{M}_\odot}}\xspace}

\def\teff{\ensuremath{T_{\mathrm{model}}}\xspace}

\newcommand{\phx}{\texttt{PHOENIX}\xspace}

\graphicspath{{./figures/eps/}{./figures/pdf_png/}}

\begin{document}

\title{Searching for Hydrogen in Type Ib Supernovae}

\author{Spencer James and E.~Baron\altaffilmark{1}}

\affil{Homer L.~Dodge Department of Physics and Astronomy, University
  of Oklahoma,\\ 440 West Brooks, Rm.~100, Norman, OK 73019-2061, USA}

 \altaffiltext{1}{Computational Cosmology Center, Lawrence Berkeley
   National Laboratory, MS 
   50F-1650, 1 Cyclotron Rd, Berkeley, CA 94720-8139 USA}

\begin{abstract}

  We present synthetic spectral fits of the typical Type Ib SN~1999dn
  and the Hydrogen Rich Ib SN~2000H using the generalized
  non-local thermodynamic equilibrium stellar atmospheres code \phx.
  We fit model spectra to five epochs of SN~1999dn ranging from ten
  days pre-maximum light to 17 days post-maximum light and the two
  earliest epochs of SN~2000H available, maximum light and six days
  post-maximum.  Our goal is to investigate the possibility of
  hydrogen in Type Ib Supernovae~(SNe Ib), specifically a feature around
  6200\AA\ which has previously been attributed to high velocity
  H$\alpha$.  In earlier work on SN~1999dn we found the most plausible
  alternative to H$\alpha$ to be a blend of Si~II and Fe~II lines
  which can be adjusted to fit by increasing the metallicity.  Our
  models are simple; they assume a powerlaw density profile with
  radius, homologous expansion, and solar compositions. The helium core
  is produced ``burning'' 4H$\rightarrow$He in order to conserve
  nucleon number.  For models with hydrogen the outer skin of the
  model consists of a shell of solar composition.  The hydrogen mass
  of the standard solar composition shell is $M_H\la$~$10^{-3}M_\odot$
  in SN~1999dn and $M_H\la$~$0.2M_\odot$ for SN~2000H.  Our
  models fit the observed spectra reasonably well, successfully
  reproducing most features including the characteristic He I
  absorptions.  The hydrogen feature in SN~1999dn is clear, but much more
  pronounced in  SN~2000H.  We discuss a possible evolutionary
  scenario that accounts for the dichotomy in the hydrogen shell mass
  between these two supernovae.

\end{abstract}


\section{Introduction}

\subsection{Supernovae Nomenclature}

Supernovae (SNe) are divided into two main subtypes, Type~I and Type~II,
based on the presence of hydrogen in their spectra. SNe of Type I are then
further subdivided into Types Ia, Ib, and Ic \citep{wh90}.  SNe~Ia  are
characterized by strong Si~II absorption near 6100\AA\ and a
characteristic sulfur `W'
shaped feature at 5400\AA. SNe~Ib  are classified by the presence
of optical He~I lines in their spectra, and SNe~Ic are noted for their lack of
helium and silicon. SNe~Ia are thought to be caused by the
thermonuclear explosion of a Chandrasekhar mass white dwarf star while SNe II,
Ib, and Ic  are caused by the core collapse of a massive star.
The general consensus is  that most SNe~Ib/c result from mass loss due to
interaction with a binary companion and that SNe~IIb also result from
binary interaction 
\citep{pod92,podnat93,nomoto_sn1bc_rev95,heger03}.

Supernovae are classified by their spectra.
 As early
spectra become more commonly available, trends in
the variation 
of certain SNe types over the course of time have emerged.  This is
apparent in 
SN~1993J which was originally classified as a SN II for its
conspicuous hydrogen but later showed helium absorptions
characteristic of a SN Ib.  SN~1993J is now classified as a 
SN IIb which is believed to be the explosion of a star that has
undergone enough mass loss to lose a large portion of its hydrogen
envelope \citep{filmatho93}.  SN~1987K \citep{f87k} and SN~2008ax
\citep{chornock08ax10} are other examples of this transitory class of
SNe.  This explains the distinct Balmer lines seen at early times which
give way to significant helium features as time progresses.  This
type of transition-like object was first suggested by \citet{wpmw87}.
It has been suggested that if mass loss from a
SN II progenitor can produce   SNe~IIb then an even larger 
mass loss would result in further stripping of the hydrogen envelope
and would be characteristic of  SNe~Ib \citep{f87k}.
This suggests that perhaps we may see hydrogen in early spectra of
 SNe Ib if there was a large enough mass of hydrogen left prior to
explosion.  \citet*{nomoto_sn1bc_rev95} also suggest several binary
mass loss scenarios that might lead to a spectroscopic sequence of SN
types from IIn$\rightarrow$IIb$\rightarrow$Ib$\rightarrow$Ic.  By
understanding the behavior of these transitory types of SNe we are
able to better extrapolate back to the state of the progenitor star.
We therefore investigate the early and late time spectra of a typical
Type Ib SN~1999dn and the early spectra of a Hydrogen Rich SNe~Ib
SN~2000H \citep{branIb02} in search of clues as to whether or not a
thin hydrogen skin exists prior to the supernova explosion.

\subsection{Prior Investigations}

The idea that there exists an underlying physical connection between
SNe~II and SNe~Ib is not a new one. As more SNe are discovered at
earlier times we see a number of these SNe~IIb, which appear to be
transition-like objects from Type II$\rightarrow$Ib.  The recent
discovery of SN~2008ax \citep{chornock08ax10} has led to more detailed
investigations into the connection between SNe~IIb and SNe~Ib, and
\citet{chornock08ax10} suggested that a full non-local thermodynamic
equilibrium (NLTE) treatment is necessary for a conclusive analysis of
the details of the hydrogen skin.

\citet{branIb02} conducted a direct analysis of a selection of SN~Ib
spectra obtained by \citet{math_sn1bc} using the supernova spectrum
synthesis code SYNOW.  They concluded that a small amount of hydrogen
is present in perhaps all  SNe~Ib spectra, specifically SN~1999dn, and
that even 
more is seen in what they term Hydrogen Rich SNe~Ib: 2000H,
1999di, and 1954A.  Others  \citep{elm_1bc06,dengetal00,parrent05bf07}
have also used SYNOW to 
investigate the presence of hydrogen  in SNe~Ib.

Because the relative simplicity of SYNOW a first principles non-local
thermodynamic equilibrium (NLTE) treatment is needed to analyze the
presence of hydrogen in detail. While a SYNOW analysis can determine
whether a particular ion is there or not and can also indicate the
relative strength and velocity extent of that species, one cannot use
SYNOW to determine quantative abundances. We use the NLTE
stellar atmospheres code \phx \citep{hbmathgesel04,hbjcam99} to
accomplish this. We perform a time-series analysis of the spectra
of SN~1999dn, a typical SN~Ib, as well as some calculations of early
time spectra of another SN~Ib, SN~2000H. \citet{benetti2000H}
originally classified SN~2000H as a SN~IIb due to strong $H_\alpha$ and
possible $H_\beta$ absorptions.

We wish to accomplish the following: (1) Produce hydrogen and
hydrogen-free fits to relevant epochs of SN~1999dn and SN~2000H; (2) Constrain the
possible ejection velocity and mass of the hydrogen skin.

\section{Methods}

\subsection{General Methods}

We generate synthetic spectra of SNe~Ib using the non-local
thermodynamic equilibrium (NLTE) stellar atmospheres code \phx
\citep[][and references therein]{hbmathgesel04,hbjcam99} The inclusion
of NLTE calculations allows us to investigate the importance of
non-thermal processes in reproducing SN~Ib spectra.  We explore the
possibility of the existence of hydrogen in SNe~Ib by adding a thin
skin of solar composition material \citep{GAS07} to a helium core
(which also has solar compositions of metals). The helium core is
constructed by beginning with material at solar composition and then
``burning'' 4H$\rightarrow$He.  We then add the hydrogen skin with
solar abundances by
replacing the abundances above a specified velocity, $v_H$, which is
varied to find the best fit
without disturbing  the helium core 
interior to $v_H$.  Our models extend out to a
maximum velocity of 21000~\kmps regardless of whether they are pure
helium cores or include a hydrogen skin.   We approximate the gamma-ray deposition
function by a constant parameter that is adjusted for each epoch. The
assumption of a sharp boundary between the hydrogen and helium layers
is not realistic, but is simple and eliminates the need to add extra
parameters.  This approximation should not significantly
effect our conclusions.

The earliest model of SN~1999dn is only  $M=$~$0.85M_\odot$ as
compared with all the other days that are  $M=$~$38M_\odot$.  This
difference in mass was necessary in order deal with the large
variation in the dynamic range of the atmosphere. Nevertheless our
models are internally consistent. The very large mass of the
progenitor is due to the straightforward extrapolation of the steep
powerlaw density profile to low radii. Here it is just a calculational
expedient and does not imply that the progenitor was anywhere close to
this massive.

\section{Results}

\subsection{SN~1999dn}

\begin{figure}
\centering
\includegraphics[width=.95\textwidth]{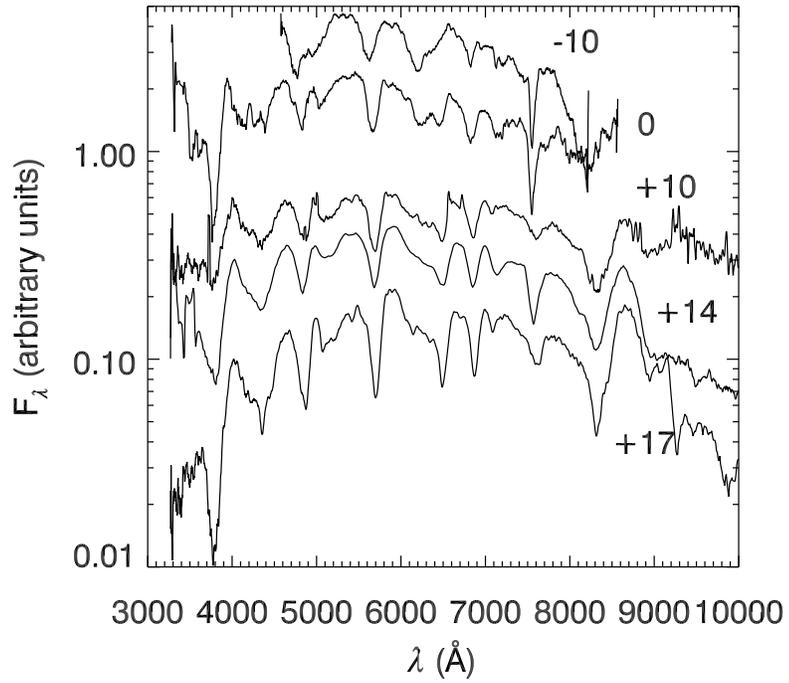}
\caption{SN~1999dn Epochs: Emphasis is placed on the 6200\AA\ 
  absorption feature that fades after maximum light. Observed spectra
  \citep{math_sn1bc,dengetal00} 
  have been smoothed with a 10 point boxcar. These spectra were all
  obtained from the Supernovae Spectrum Repository (SUSPECT)
  http://suspect.nhn.ou.edu maintained by the University of Oklahoma.}
\label{fig:fig1}
\end{figure}

Fig.~\ref{fig:fig1} shows the observed spectra for several  epochs of
SN~1999dn \citep{math_sn1bc,dengetal00}.  While we seek the overall best fit to
the spectra,  emphasis is placed on the absorption feature
around 6200\AA\ which fades after maximum light.  The observed spectra
are smoothed with a 10 point boxcar average in order to remove some
noise.

\begin{figure}
\centering
\includegraphics[width=.95\textwidth]{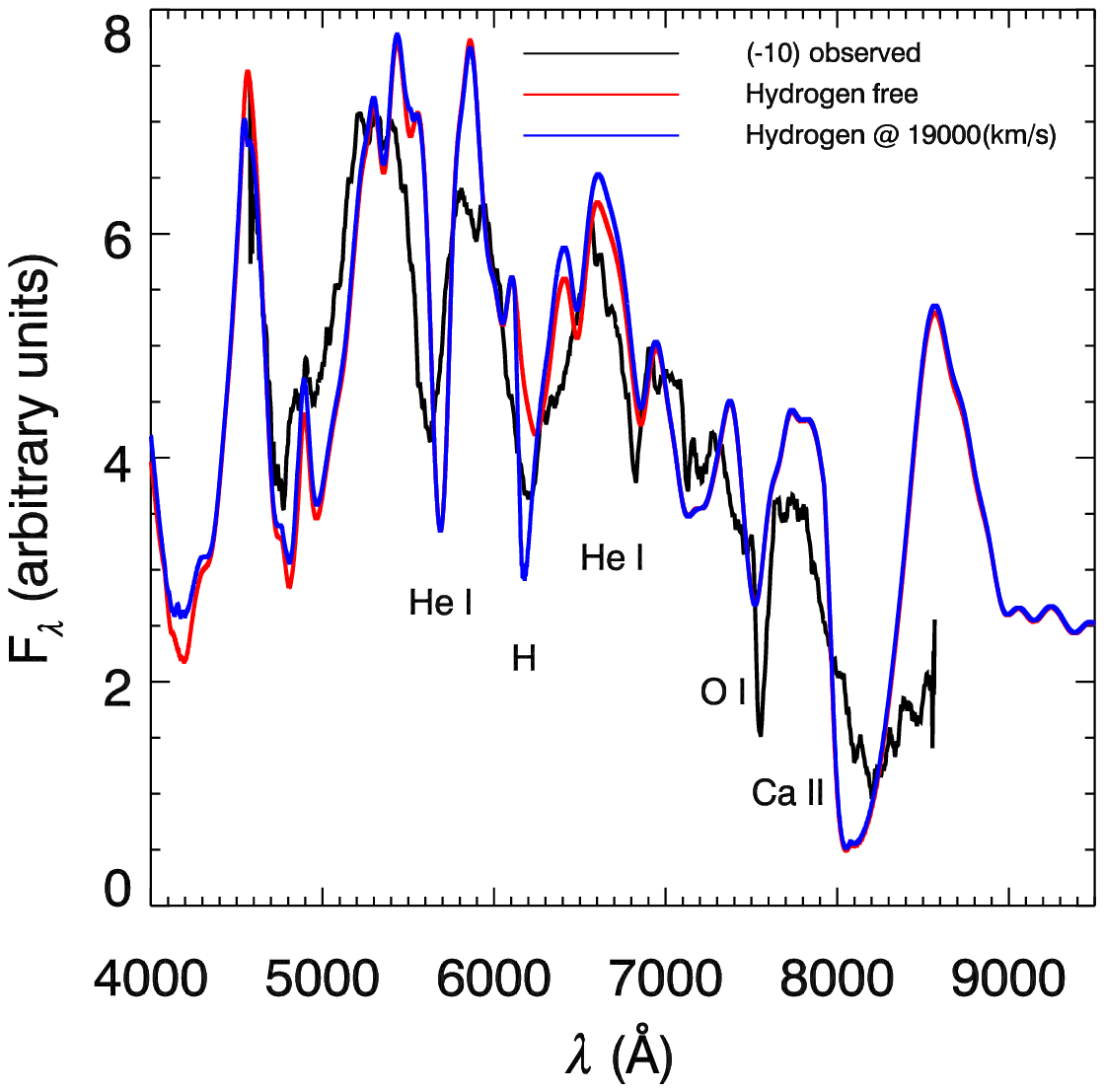}
\caption{SN~1999dn (-10) Observed spectrum \citep{dengetal00}
  smoothed with a 10 point boxcar compared to synthetic spectra with
  and without hydrogen at $v_H = 19000$~\kmps.}
\label{fig:fig2}
\end{figure}

\begin{figure}
\centering
\includegraphics[width=.95\textwidth]{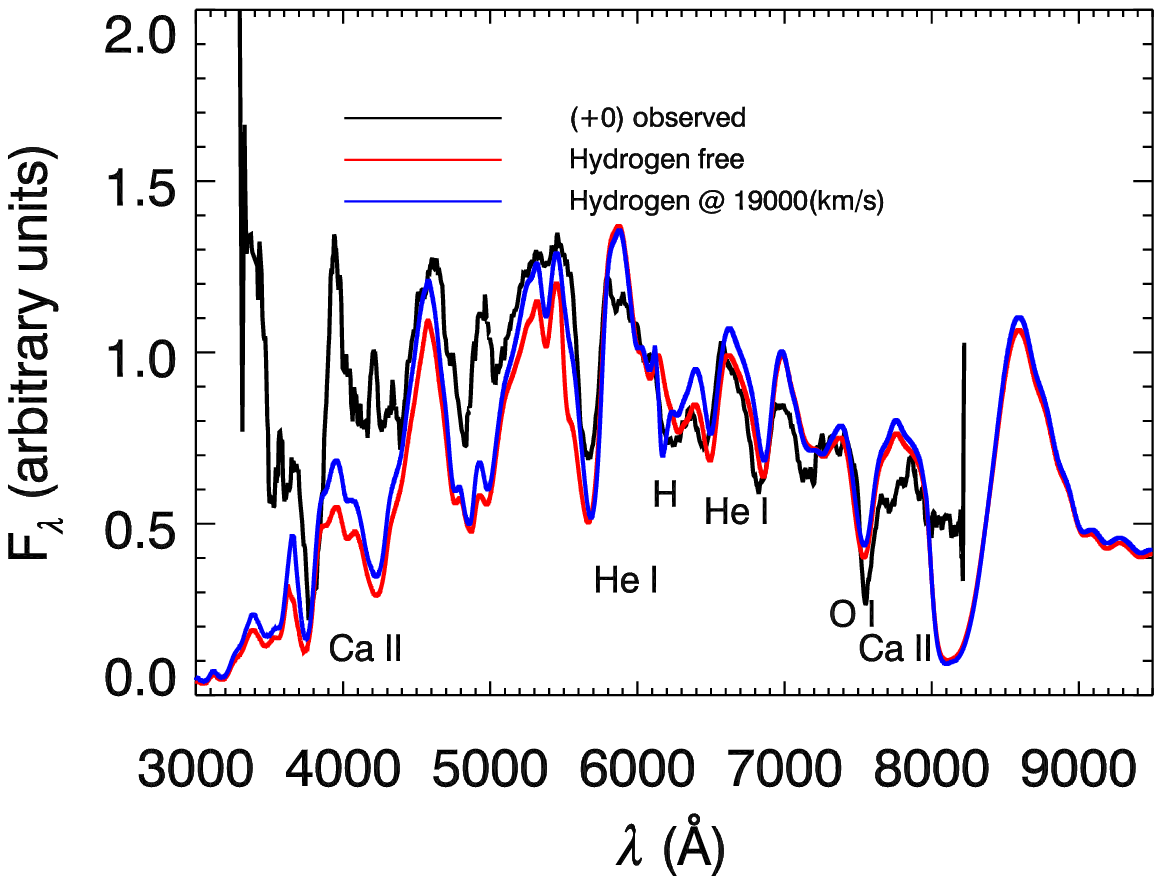}
\caption{SN~1999dn (0) Observed spectrum \citep{dengetal00} smoothed
  with a 10 point boxcar compared to synthetic spectra with
  and without hydrogen at $v_H = 19000$~\kmps.}
\label{fig:fig3}
\end{figure}

\begin{figure}
\centering
\includegraphics[width=.95\textwidth]{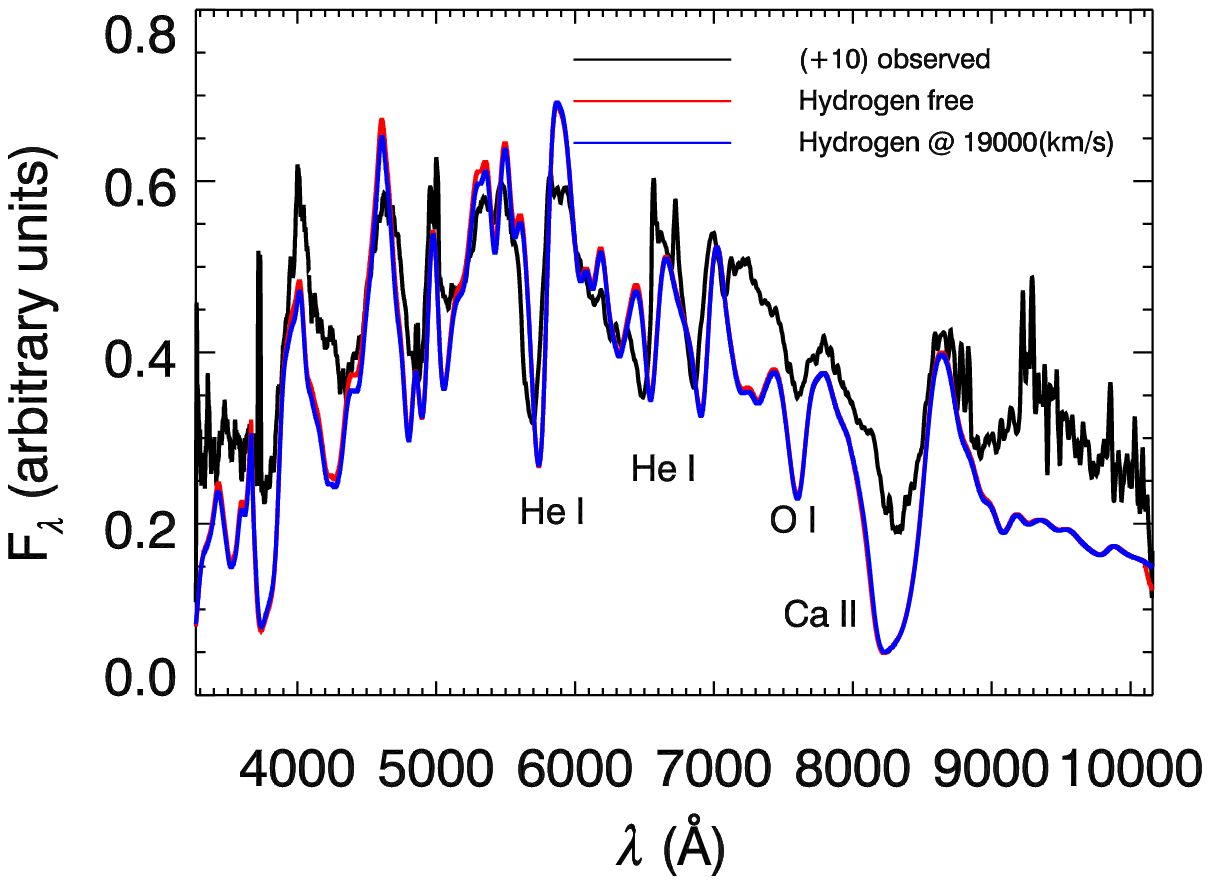}
\caption{SN~1999dn (+10) Observed spectrum \citep{math_sn1bc}
  smoothed with a 10 point boxcar compared to synthetic spectra with
  and without hydrogen at $v_H = 19000$~\kmps.}
\label{fig:fig4}
\end{figure}

\begin{figure}
\centering
\includegraphics[width=.95\textwidth]{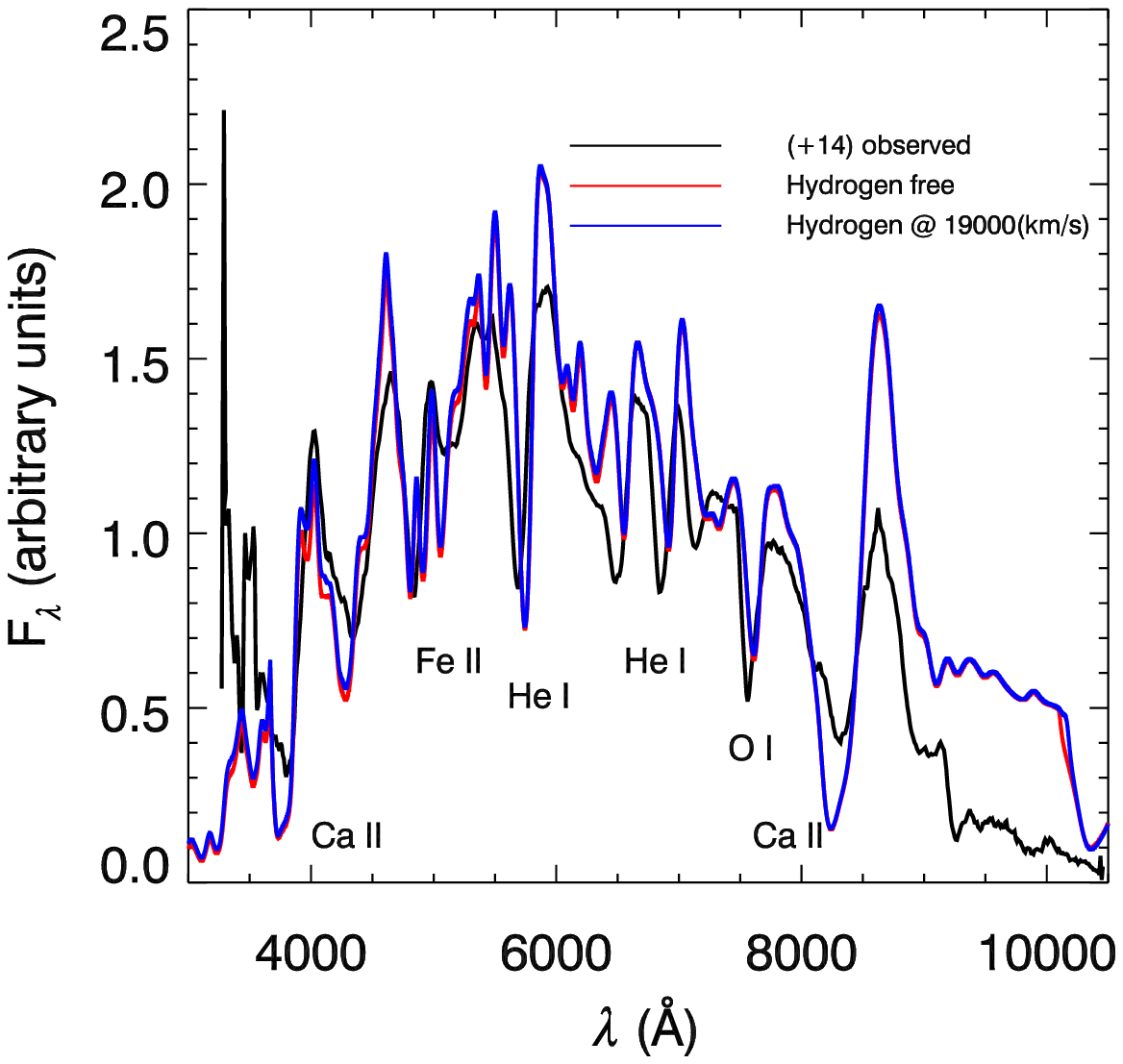}
\caption{SN~1999dn (+14) Observed spectrum \citep{dengetal00} smoothed
  with a 8 point boxcar compared to synthetic spectra with
  and without hydrogen at $v_H = 19000$~\kmps.}
\label{fig:fig5}
\end{figure}

\begin{figure}
\centering
\includegraphics[width=.95\textwidth]{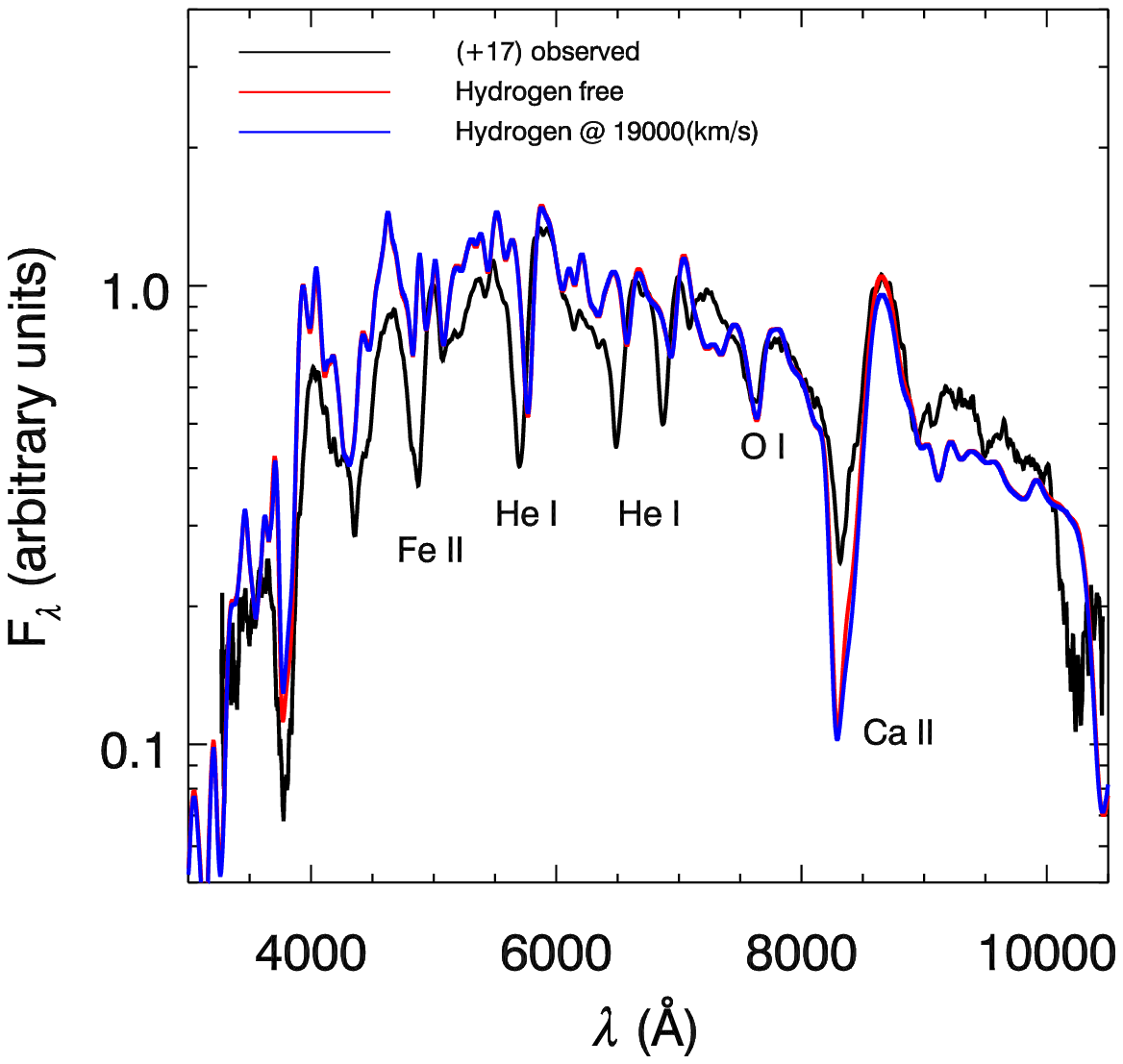}
\caption{SN~1999dn (+17) Observed spectrum \citep{math_sn1bc} smoothed
  with a 10 point boxcar compared to synthetic spectra with
  and without hydrogen at $v_H = 19000$~\kmps.}
\label{fig:fig6}
\end{figure}

\begin{figure}
\centering
\includegraphics[width=.95\textwidth]{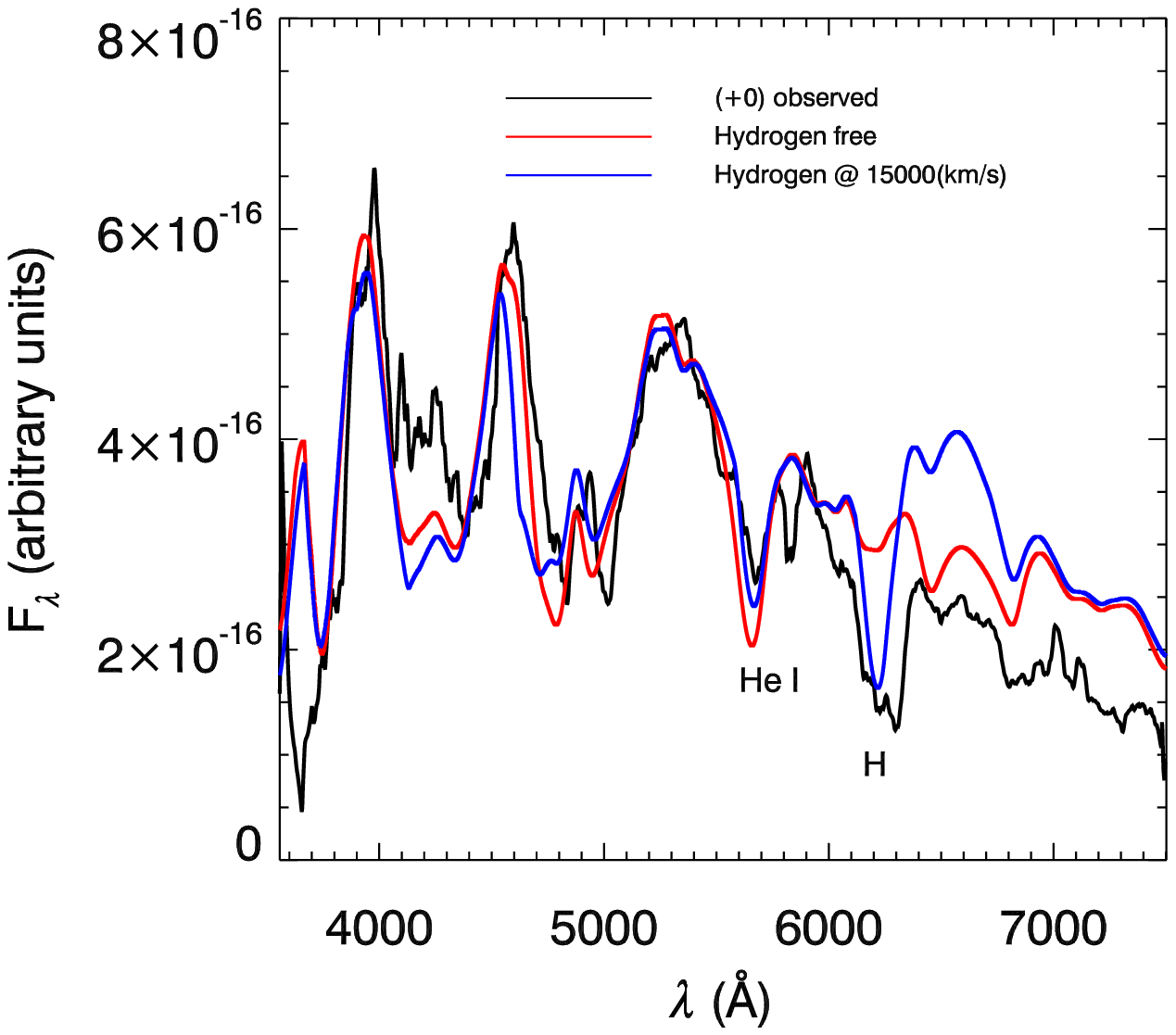}
\caption{SN~2000H (0) Observed spectrum \citep{Asiago09} smoothed with a 20 point
  boxcar compared to synthetic spectra with
  and without hydrogen at $v_H = 15000$~\kmps.}
\label{fig:fig7}
\end{figure}

\begin{figure}
\centering
\includegraphics[width=.95\textwidth]{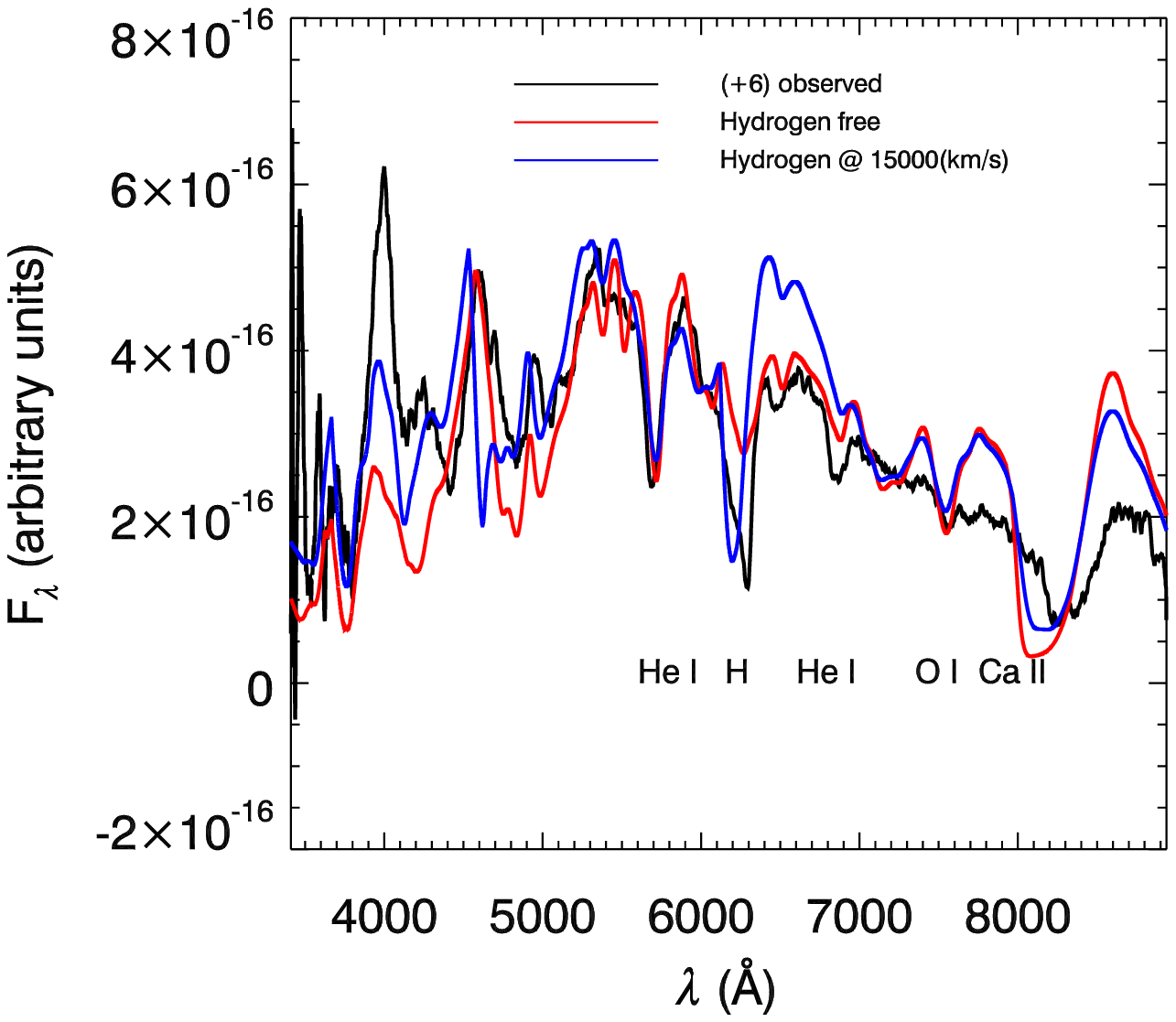}
\caption{SN~2000H (+6) Observed spectrum \citep{Asiago09} smoothed
  with a 20 point boxcar compared to synthetic spectra with
  and without hydrogen at $v_H = 15000$~\kmps.}
\label{fig:fig8}
\end{figure}

We explored variations in multiple parameters for each epoch adjusting
the total bolometric luminosity in the observer's frame (specified by
a temperature, \teff), the photospheric velocity, the metallicity, the
density profile,
gamma-ray deposition, and hydrogen mass.  By
varying these parameters 
we have arrived at a set of parameters which we consider the best 
and we follow that model through each epoch.  The relevant
parameters for SN~1999dn are given in Table~\ref{tab:tab1} and in
Table~\ref{tab:tab2} for SN~2000H. The results of the NLTE \phx
calculations are shown in
Figs.~\ref{fig:fig2}--\ref{fig:fig8}. 
We treat the following species in NLTE: H~I,
He~I-II, C~I-III, N~I-III, O~I-III, Ne~I, Na~I-II, Mg~I-III, Si~I-III,
Ca~I-III, and Fe~I-III. In all the
spectra we successfully reproduce the characteristic He I absorptions 
at 5876\AA, 6678\AA, and 7065\AA.

Fig.~\ref{fig:fig2} shows the the early epoch
Aug.~21 for SN~1999dn \citep[-10 days]{math_sn1bc} the synthetic spectrum with hydrogen at 19000 \kmps fits
the 6200\AA\ feature better than a pure helium core.  In the maximum light
spectrum,
Aug.~31, the hydrogen feature at 19000~\kmps
does not properly blend with the feature just to the red 
(Fig.~\ref{fig:fig3}), but this is likely due to our assumption of a
sharp abundance boundary.  The detached-like profile (blueshifted
absorption with flat topped emission) is a
consequence of 
the abrupt cutoff of our model.  The hydrogen velocity is 19000~\kmps
and the outer velocity layer of our model is 21000~\kmps.  A smoother
variation in the abundances would reduce 
the sharp cutoff that leads to the detached profile.

By Sept.~10 the hydrogen feature has faded in both the observed
spectrum as well as the synthetic spectra as seen in
Fig.~\ref{fig:fig4}.  This is reinforced by the lack of a feature in
the later days of the observed spectra.

In Fig.~\ref{fig:fig5} and Fig.~\ref{fig:fig6}, on Sept.~14 and Sept.~17
respectively, we find that the
inclusion of hydrogen has little if any effect on the spectrum and so
conclude that the thin skin of hydrogen is no longer visible due to
geometric dilution.

The He I absorptions in many of the spectra appear
to be slightly to the red of the observed absorptions.  This is
probably due to the
assumption of a constant gamma-ray deposition parameter.  By increasing
the amount of gamma-ray deposition we increase the non-thermal
components of the spectrum which increases the depth of the He I
absorptions and also the speed of highly non-thermal lines.  A more
thorough treatment of the nickel distribution would likely improve
this, but examining the exact amount of nickel mixing is beyond the
scope of the present work.

\subsection{SN~2000H}

The earliest spectrum available of SN~2000H is at maximum light, Apr
21 2000 \citep{Asiago09}. 
Fig.~\ref{fig:fig7} shows
a very deep and wide absorption trough in the region of $H_\alpha$
absorption.  Part of the absorption feature could be due to blending
with other elements due to circumstellar interaction.  Towards the end
of this section we discuss the narrow Na D absorption near 5800\AA\
and the information we can gather about the circumstellar medium (CSM)
from it.

Part of the line profile could be due to a low velocity CSM region
  but the depth of the line would suggest that we should also see
  evidence of $H_\beta$, which is lacking.
This supernova is highly reddened.  We adopt a reddening value of
$E(B-V)=0.61$ \citep{benetti2000H} and a recession velocity of
$v_{rec}=3945\ \kmps$ taken from the
NED\footnote{http://nedwww.ipac.caltech.edu/} database.

We find that the spectrum of SN~2000H at +6 days post max light
\citep{Asiago09} can
also be well reproduced by adding hydrogen to our models as is shown in
Fig.~\ref{fig:fig8}.  We find that a hydrogen velocity of
$v_H=$~15000~\kmps is necessary to fit the hydrogen feature around
6200\AA.  The amount of hydrogen included in the 2000H
model ($M_H\la$~$0.2M_\odot$) is significantly greater than in our
SN~1999dn models,
($M_H\la$~$10^{-3}M_\odot$), which is consistent with the
classification of a Hydrogen Rich SN~Ib by \citet{branIb02}.  This is
necessary to fit the hydrogen feature that persists to as late as +6
days post-max light. A larger amount of hydrogen suggests that
SN~2000H is closer to the Type IIb subtype due to a lower degree of
envelope stripping than in SN~1999dn
\citep{benetti2000H,blondin07a,chornock08ax10}. 

\begin{figure}
\centering
\includegraphics[width=.95\textwidth]{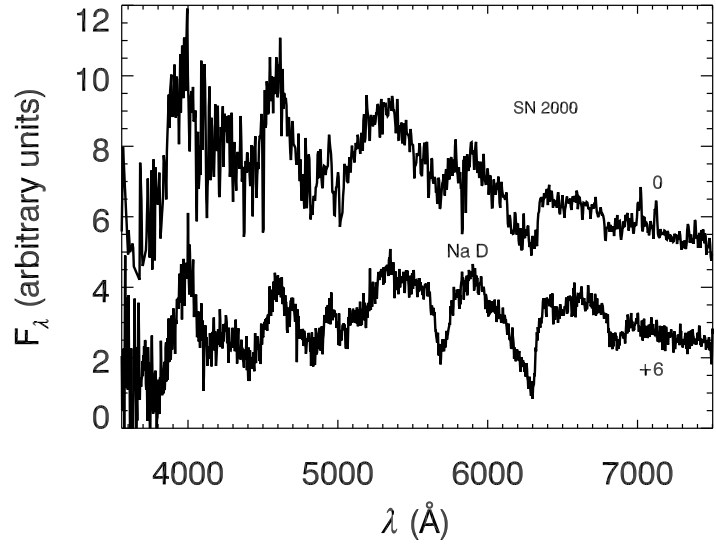}
\caption{SN~2000H epochs \citep{Asiago09}  without any smoothing.}
\label{fig:fig9}
\end{figure}

\begin{figure}
\centering
\includegraphics[width=.95\textwidth]{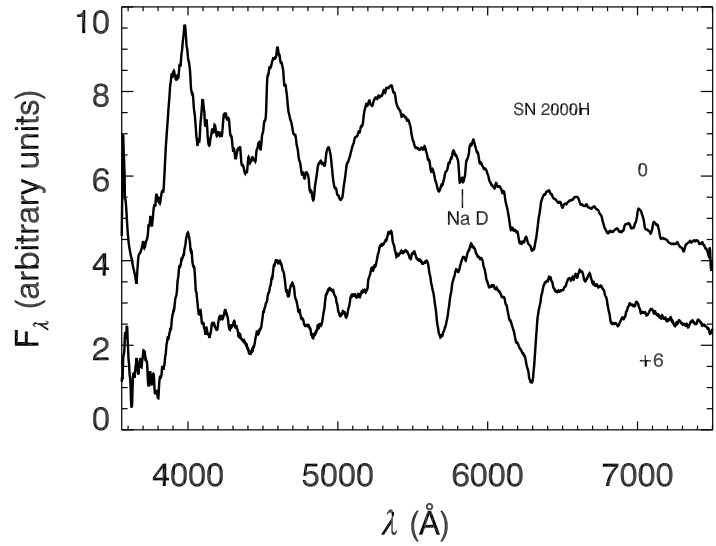}
\caption{SN~2000H epochs \citep{Asiago09} smoothed with a 20 point boxcar.}
\label{fig:fig10}
\end{figure}

\begin{figure}
\centering
\includegraphics[width=.95\textwidth]{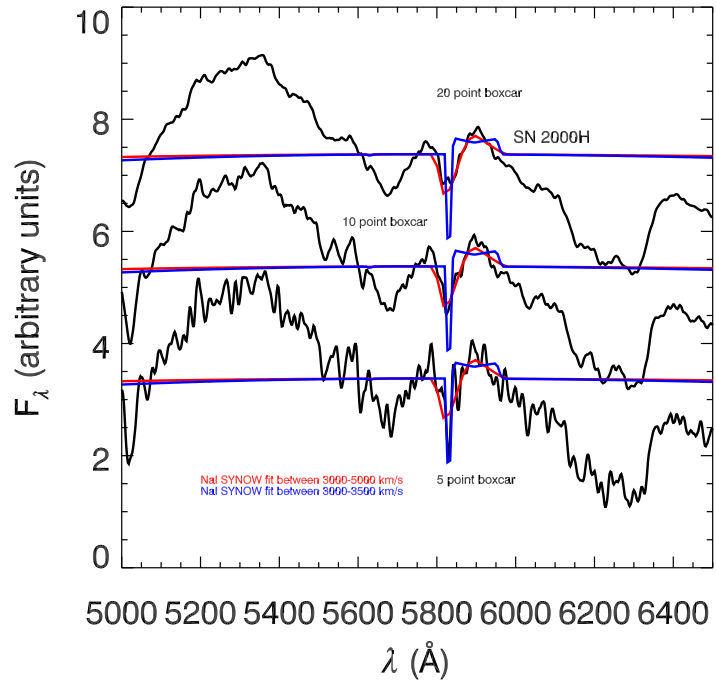}
\caption{SN~2000H (0) \citep{Asiago09} with a 20, 10, and 5 point
  boxcar plotted with a SYNOW spectrum of Na I between 3000--5000
  \kmps and detached from the photosphere but constrained to a 
  velocity between 3000-3500 \kmps.}
\label{fig:fig11}
\end{figure}

Figs.~\ref{fig:fig9}--\ref{fig:fig10} show two of the observed
spectra of SN~2000H both with and without a 20 point boxcar smoothing.
The narrow Na D absorption
around 5800\AA\ appears to fade quickly since it
is not seen in the spectrum taken 6 days later.  In
Fig~\ref{fig:fig11} we present a comparison of a SYNOW spectrum of
only Na I constrained to 3000--5000 \kmps as well as the case where
the Na I is detached from the photosphere but constrained to
3000--3500 \kmps with the maximum light
spectrum of SN~2000H with 20, 10, and 5 point boxcar smoothing.  The
Na I and the 20 point boxcar spectrum look like the best fit, but since
the SYNOW spectrum appears too broad to fit the 5 point boxcar the
detached spectrum is a better fit in that case. In either case one
finds a characteristic velocity of the Na D line to be about 3000~\kmps.

We believe the Na D feature to be formed from the surrounding
CSM.   The SYNOW velocity is large compared to typical wind velocities.
\citet{YWL10} found a wind velocity for different SN~Ib/c
progenitors to range from $\sim200~\kmps$ to $\sim2400~\kmps$ which at the
upper limit could explain the high velocity we find for the Na D
feature from the CSM, however with a hydrogen shell of $\sim
0.1$~\msol, the wind is more likely to be of order 500~\kmps. The high velocity could also be due to radiative
acceleration at shock breakout. We find that the total energy required
to accelerate this mass to be $E_{\mathrm{kin}} = 2\times 10^{47}\ \mathrm{ergs}
  ({\rho}/{10^{-15}\ \mathrm{g\ cm}^{-3}})$ where the density is
  obtained from the SYNOW fit assuming an ionization fraction of
  $10^{-3}$. 

\section{Discussion}

We find that the results from the \phx calculations suggest a hydrogen
mass of $M_H\la$~$10^{-3}M_\odot$ for SN~1999dn. This corresponds to a hydrogen
velocity $v_H\ga$~19000~\kmps. \citet{tanaka08D09} placed a limit of
$M_H\la$~$5\times 10^{-4}M_\odot$ on the hydrogen mass in SN~2008D and a
limit on the Doppler velocity of approximately $v_H\approx$~18500
\kmps. Others have also suggested that hydrogen is responsible for this
feature around 6200\AA. \citet{dengetal00} suggested a hydrogen
velocity for SN~1999dn of 19000 \kmps on Aug.~21, and 18000 \kmps on
Aug.~31, and that the feature becomes blended and then taken over by C
II in later days. \citet{ketchum08} investigated C II, Si II, Fe II,
and Ne I as possible candidates as well but found that without
increasing the metallicity and abundances beyond physically reasonable amounts
the 6200\AA\ feature was not easily reproduced in early epochs. This
seems to be consistent with our calculations, reinforcing the
the existence of hydrogen in early spectra of SNe Ib.

We also find that the results from the \phx calculations suggest a hydrogen
mass of $M_H\la$~$0.2 M_\odot$ for SN~2000H.  This corresponds to
a hydrogen velocity in our models of $v_H\approx$~15000~\kmps.  This
mass estimate is likely an upper limit since we assume a single
powerlaw density profile for both the helium core and hydrogen
shell. The width 
of the 6200\AA\ feature early on is likely due to some blending with some
circumstellar material since it disappears in the spectrum 6 days
later and the hydrogen feature is more clear.  We know that circumstellar
interaction was taking place at maximum light from the presence of the
narrow Na D feature that quickly fades.  The larger amount of
hydrogen is more characteristic of that seen in a SN~IIb such as
SN~1993J or SN~2008ax with mass estimates on the order of
$M_H\ga$~$0.1 M_\odot$ \citep{wheel94I94,chornock08ax10}. 

The possibility of a hydrogen skin existing prior to the
supernova explosion is interesting. \citet{pod92} first discussed the
evolution of binary systems that would lead to mass stripping.
\citet{nomoto_sn1bc_rev95} discuss a 
continuum of supernovae types from
IIn$\rightarrow$IIb$\rightarrow$Ib$\rightarrow$Ic which would be
consistent with the idea of a varying degree of hydrogen mass loss due to a
strong wind or binary interaction in a non-conservative mass
transfer scenario. The energy released in a binary merger would
be sufficient to eject some of the envelope.   \citet{wheel94I94}
suggested that if hydrogen is present in  SNe Ib then
the amount of hydrogen in  SNe Ib is less than that found in the
Type IIb SN~1993J, which was about
$0.1-0.5$~$M_\odot$ \citep{filmatho93}.
This is consistent with our results of SN~1999dn but suggests that
SN~2000H is more accurately classified as a SN~IIb.
\citet{chev_frans06} also suggest that there is a continuous distribution
of SNe between SNe IIb where the hydrogen is clearly present and 
SNe Ib where the hydrogen is quite weak.  This continuous distribution
is likely to be the explanation for the sequence of SNe between type
IIb and Ib and the varied amounts of envelope stripping are a result
of binary interaction and/or strong stellar winds. SN~1993J is known to
have been in a binary and as a result had some of its hydrogen
envelope lost due to the merging of the two stars in the binary
\citep{podnat93,nomoto_sn1bc_rev95}.

Recently \citet{YWL10} studied binary progenitors of SNe~Ib/c
including the effects of rotation and various reduced wind-loss
formulations. They find that even in the case of Case A and B mass
transfer that the mass transfer is non-conservative and that a limited
range of helium cores ($3.5 \la M \la 4.5$~\msol, for solar
metallicity and $3.5 \la M \la 8 $~\msol, for $Z_\odot/20$) that the
wind loss in the helium WR stage is low enough that a hydrogen skin of
the order of $M_H \sim 10^{-2} - 10^{-3}$ can be retained.

Even though there is a rather continuous distinction between SNe~IIb
and hydrogen rich SNe~Ib, the total  hydrogen mass in the envelope is
rather similar $M_H \sim 0.1-0.3$~\msol. These objects are likely from binary
progenitors with stable Case C mass transfer, which leads the donor to
shrink below the Roche limit when the envelope reaches a critical mass
of about 0.3~\msol \citep{pod92,podnat93}. Further reduction of the
envelope mass then follows via ablation of the hydrogen envelope from
the wind of the secondary \citep{pod92,podnat93}. However, studies of SNe~Ib
suggest that many if not most have a much thinner skin of hydrogen
similar to what we find here for SN~1999dn
\citep{branIb02,chornock08ax10}. This could likely be explained by a
binary scenario that leads to a common envelope that is ejected from
the system \citep{pod92}. As described in \citet{pod92} this will lead
to either a cataclysmic variable (if the core mass is less than
1.4~\msol) or to a low mass helium core, which would explode as a
SN~Ib, but in the common envelope environment it would not be hard to
keep a thin hydrogen skin. If this is a  common channel for SNe Ib
formation than one would expect many to have thin hydrogen envelopes
which will be visible in early spectra. Since massive helium cores
will blow strong winds that would quickly remove a $10^{-3}$~\msol
envelope this lends support that many SNe~Ib progenitors
are the result of relatively low mass helium cores.

\section{Conclusion}

Our results indicate that hydrogen does exist in the Type Ib SN~1999dn
and SN~2000H with an upper mass limit of $M_H\la$~$10^{-3}M_\odot$ and $M_H\la$~$10^{-1}M_\odot$ respectively.  The
existence of such a small amount of hydrogen is to be expected as
suggested by the continuous spectrum of SN types from II to Ic.
Further analysis of a more robust sample of SN Ib is needed for a
definitive claim, but the existence of a hydrogen skin in  SNe Ib
is seen in the observed spectra as well as supported by synthetic calculations.  This
implies that the typing of core-collapse SNe is more useful as a means
of determining the state of the ejecta at a given epoch as opposed to
inferring
the state of the progenitor at the time of explosion.  While it is
still useful to determine a SN type we need to set a uniform epoch,
such as maximum light or later, in order to be consistent and
conclusive when it comes to classification.

\acknowledgements We thank Phillip Podsiadlowski for extensive
discussions and tutelage on the nature of binary interactions in
stripped envelope supernovae. This work was supported in part  
NSF grant AST-0707704, and US DOE Grant
DE-FG02-07ER41517.    This research used resources of the National Energy
Research Scientific Computing Center (NERSC), which is supported by the Office
of Science of the U.S.  Department of Energy under Contract No.
DE-AC02-05CH11231.

\bibliography{refs,baron,sn1bc,sn1a,sn87a,snii,sn93j,stars,rte,cosmology,gals,agn,atomdata,local,crossrefs}

\begin{deluxetable}{ccccc}
\tablecolumns{5}
\tablewidth{0pc}
\tablecaption{Model Parameters: 1999dn \label{tab:tab1}}
\tablehead{
\colhead{Epoch}  & \colhead{\teff (K)}  &
    \colhead{$v_{0}$ (\kmps)}  &
    \colhead{$v_{H}$ (\kmps)} &
    \colhead{${M_{H}}$ (\msol)}  
}
\startdata
{-10}    &{6000}  &{11000}  &{19000}  &{$2.95\times 10^{-4}$} \\
{0}      &{5250}  &{10000}  &{19000}  &{$2.94\times 10^{-3}$} \\
{10}     &{5400}  &{7000}  &{19000}  &{$2.63\times 10^{-4}$} \\
{14}     &{5000}  &{7000}  &{19000}  &{$2.63\times 10^{-4}$} \\
{17}     &{4600}  &{7000}  &{19000}  &{$3.38\times 10^{-4}$} \\
\enddata
\tablecomments{Relevant Model Parameters for SN~1999dn}
\end{deluxetable}

\begin{deluxetable}{ccccc}
\tablecolumns{5}
\tablewidth{0pc}
\tablecaption{Model Parameters: 2000H \label{tab:tab2}}
\tablehead{
\colhead{Epoch}  & \colhead{\teff (K)}  &
    \colhead{$v_{0}$ (\kmps)}  &
    \colhead{$v_{H}$ (\kmps)} &
    \colhead{${M_{H}}$ (\msol)}  
}
\startdata
{0}    &{6000}  &{11000}  &{15000}  &{0.124} \\
{+6}    &{5600}  &{10000}  &{15000}  &{0.125} \\
\enddata
\tablecomments{Relevant Model Parameters for SN~2000H}
\end{deluxetable}

\end{document}